\documentclass[preprint2]{aastex}
\usepackage{url}
\begin{document}

\title{\large{\rm{ANCHORS FOR THE COSMIC DISTANCE SCALE: \\ THE CEPHEIDS U SGR, CF CAS AND CEab CAS}}}

\author{\sc \small D. Majaess$^{1,2}$, G. Carraro$^3$, C. Moni Bidin$^4$, C. Bonatto$^5$, L. Berdnikov$^{6,7}$, D. Balam$^8$, M. Moyano$^4$, L. Gallo$^1$, D. Turner$^1$, D. Lane$^1$, W. Gieren$^9$, J. Borissova$^{10}$, V. Kovtyukh$^{11,12}$, Y. Beletsky$^{13}$}

\affil{$^1${\footnotesize Saint Mary's University, Halifax, Nova Scotia, Canada.}}
\affil{$^2${\footnotesize Mount Saint Vincent University, Halifax, Nova Scotia, Canada.}}
\affil{$^3${\footnotesize European Southern Observatory, Avda Alonso de Cordova, 3107, Casilla 19001, Santiago, Chile.}}
\affil{$^4${\footnotesize Instituto de Astronom\'ia, Universidad Cat\'olica del Norte, Av.~Angamos 0610, Antofagasta, Chile.}}
\affil{$^5${\footnotesize Departamento de Astronomia, Universidade Federal do Rio Grande do Sul, Av.~Bento Gonalves 9500 Porto Alegre
91501-970, RS, Brazil.}}
\affil{$^6${\footnotesize Moscow M V Lomonosov State University, Sternberg Astronomical Institute, Moscow 119992, Russia.}}
\affil{$^7${\footnotesize Isaac Newton Institute of Chile, Moscow Branch, Universitetskij Pr. 13, Moscow 119992, Russia.}}
\affil{$^8${\footnotesize Dominion Astrophysical Observatory, Victoria, British Columbia, Canada.}}
\affil{$^9${\footnotesize Departamento de Astronom\'ia, Universidad de Concepci\'on, Casilla 160-C, Concepci\'on, Chile.}}
\affil{$^{10}${\footnotesize Departamento de F\'isica y Astronom\'ia, Facultad de Ciencias, Universidad de Valpara\'iso, Av.~Gran Breta\~na 1111, Valpara\'iso, Chile.}}
\affil{$^{11}${\footnotesize Astronomical Observatory, Odessa National University, Odessa, Ukraine.}}
\affil{$^{12}${\footnotesize Isaac Newton Institute of Chile, Odessa Branch, T. G. Shevkenko Park, 65014 Odessa, Ukraine.}}
\affil{$^{13}${\footnotesize Las Campanas Observatory, Carnegie Institution of Washington, Colina el Pino, Casilla 601 La Serena, Chile.}}
\email{dmajaess@cygnus.smu.ca}

\begin{abstract}
New and existing X-ray, $UBVJHK_sW_{(1-4)}$, and spectroscopic observations were analyzed to constrain fundamental parameters for M25, NGC~7790, and dust along their sight-lines. The star clusters are of particular importance given they host the classical Cepheids U Sgr, CF Cas, and the visual binary Cepheids CEa and CEb Cas. Precise results from the multiband analysis, in tandem with a comprehensive determination of the Cepheids' period evolution ($dP/dt$) from $\sim$140 years of observations, helped resolve concerns raised regarding the clusters and their key Cepheid constituents.  Specifically, distances derived for members of M25 and NGC 7790 are $630\pm25$ pc and $3.40\pm0.15$ kpc, respectively.
\end{abstract}

\section{{\rm \footnotesize INTRODUCTION}}
The latest suite of cosmological parameters deduced from the Planck satellite has renewed concerns regarding the Cepheid distance scale and the standard $\Lambda$CDM model \citep[e.g.,][see also \citealt{kr12} and \citealt{pe13}]{ni13}. The Hubble constant cited by the \citet{pl13} differs by $\sim5$\% from that determined by the S$H_0$ES and Carnegie Hubble projects \citep{ri11,fr12}, whereby the latter efforts rely partly on Cepheids for determining $H_0$ \citep[see also][and references therein]{tr12}. The discrepancy is sufficient to hinder efforts to constrain the nature of dark energy \citep[for which $\sigma_{w}\sim2\sigma_{H_0}$,][]{mr09}, and could be linked to a systematic offset in the Cepheid calibration, or unreliable photometry for remote extragalactic Cepheids owing to crowding effects \citep[][see also Table 2 in \citealt{ri11}]{mo04,ma10,ma11b}. Research continues on the aforementioned topics irrespective of the discrepancy, namely to resolve the putative degeneracy between the impact of crowding and metallicity on Cepheid parameters \citep[][and references therein]{ma11b,ma13,ch12}, and to strengthen the Galactic Cepheid calibration \citep[e.g.,][]{ma12,an13}. That calibration is likewise important for defining the metallicity gradient of the Milky Way \citep{lu11}, benchmarking standard candles \citep{mat12,ku12}, and constraining the behaviour of intermediate mass stars \citep[][and references therein]{ne12,bo13}.  Indeed, the Galactic calibration may serve as the basis for the cosmic distance scale in concert with LMC \citep{so08} and NGC 4258\footnote{The \citet{hu13} maser distance for NGC 4258 is $\sim5$\% larger than the original \citet{he99} result. The new distance can be reconciled with the Cepheid photometry for NGC 4258 \citep{ma06} if the crowding bias endemic to the latter is considered \citep[Fig.~4 in][and see also \citealt{bo08,bo10}, \citealt{ma10}, \citealt{br11}]{ma09}.} Cepheids. 

\begin{figure*}[!t]
\begin{center}
\includegraphics[width=8.5cm]{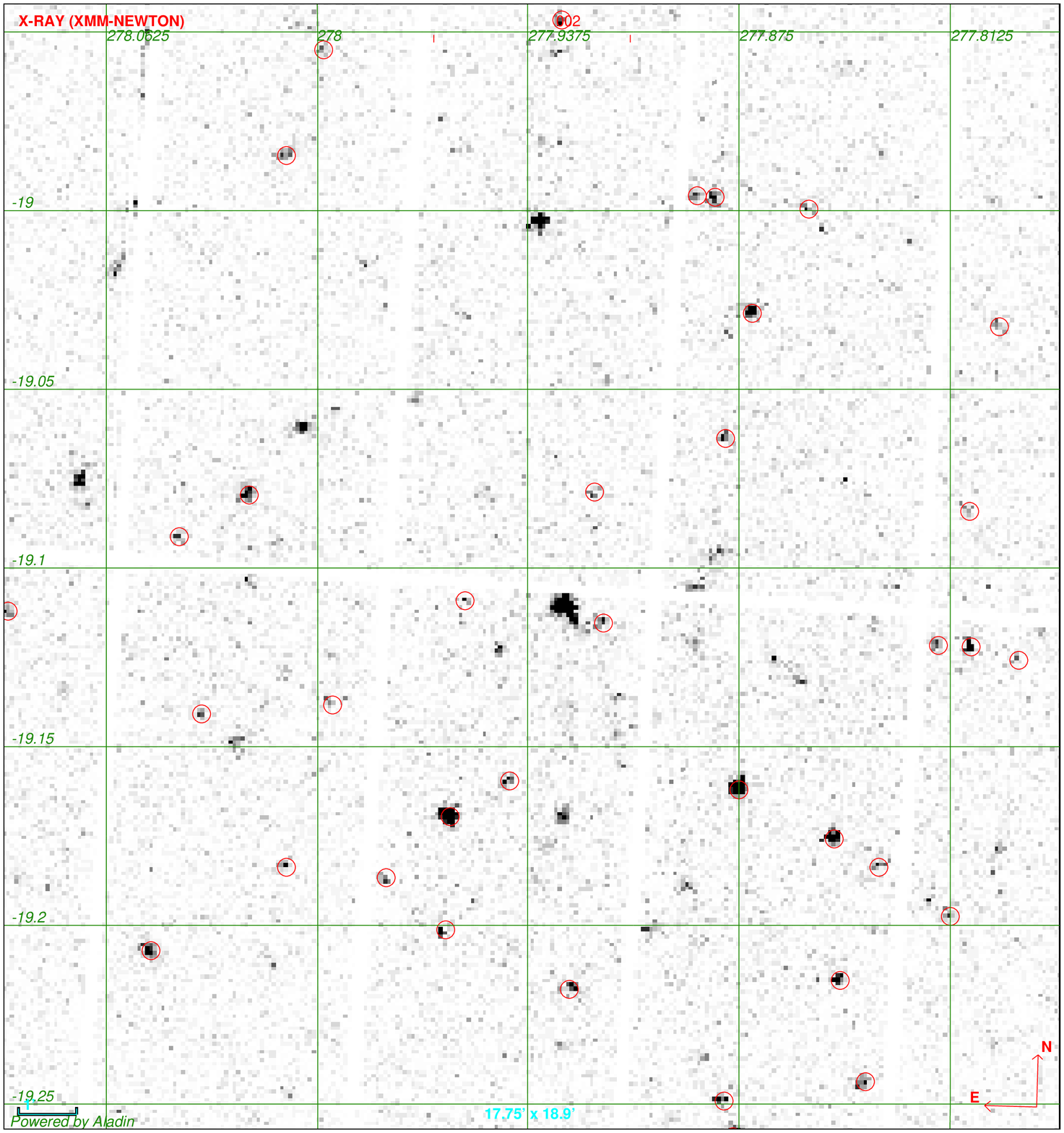} 
\includegraphics[width=8.5cm]{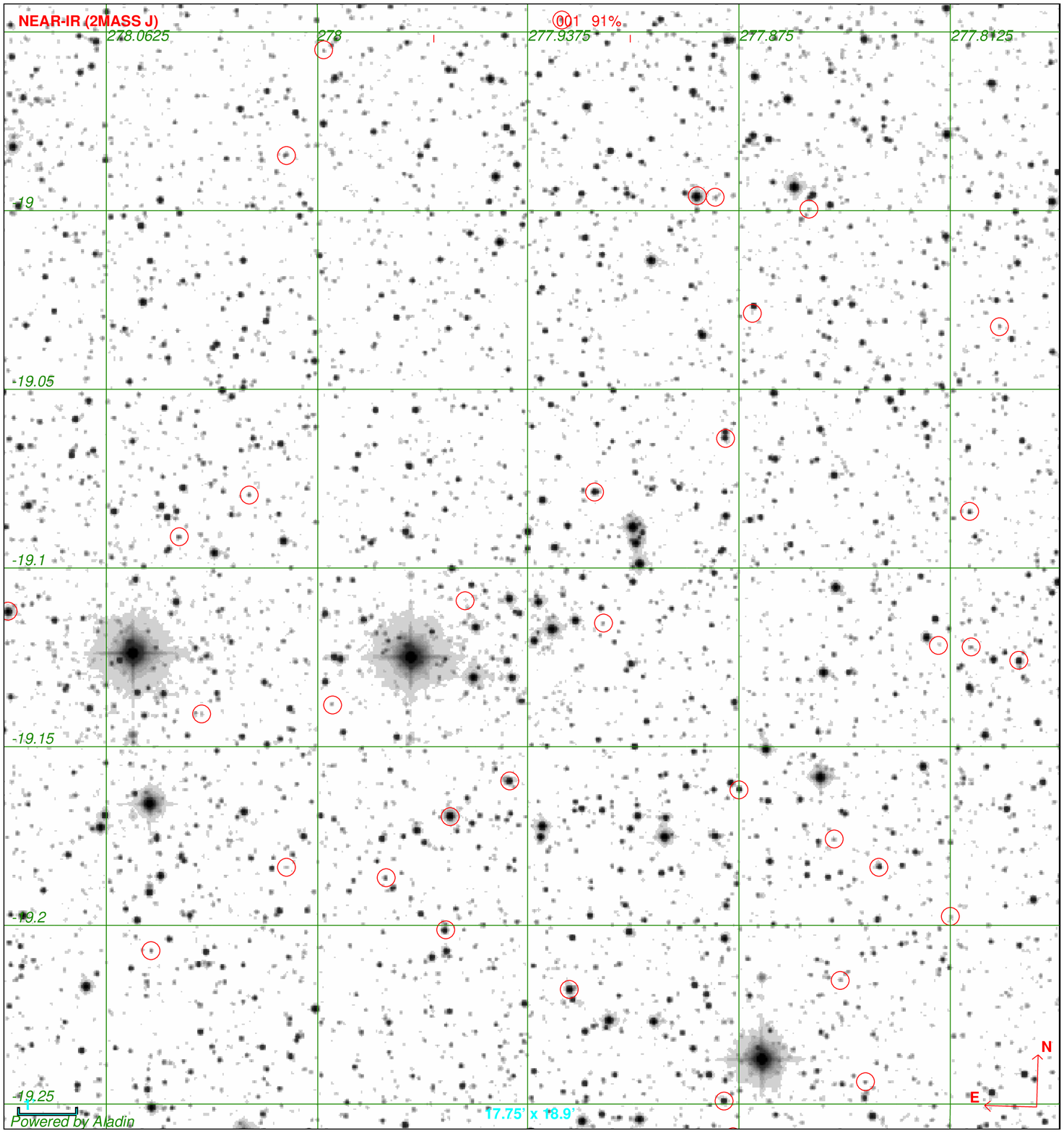} 
\caption{\small{XMM-Newton X-ray data (ObsID:040472) were used to help segregate fainter cluster members from field stars, thus facilitating the application of an isochrone and reducing uncertainties associated with the cluster distance derived \citep[e.g.,][]{ma12,ca13}.  Left, a cropped XMM-Newton image of the field encompassing M25, while a near-infrared 2MASS $J$-band image is shown on the right.  Red circles are 2MASS (AAA) sources that exhibit X-ray emission.}}
\label{fig-xmmfield}
\end{center}
\end{figure*}

Improving the Cepheid calibration by deriving solid parameters for cluster Cepheids using X-ray, $UBVJHK_sW_{(1-4)}$, and spectroscopic  observations is the principal impetus of this analysis. In particular, the star clusters M25 and NGC 7790 are examined given they host the classical Cepheids U Sgr, CF Cas, and the visual binary Cepheids CEab Cas \citep[][and references therein]{be90,sz03}. P.~Doig suggested that U Sgr may be a member\footnote{\citet{an13} hinted that Y Sgr could likewise be a member of M25.} of M25 \citep{san60}, and O.~Eggen was among the first to propose a Cepheid connection with NGC 7790 \citep{san58}. Yet generations after those discoveries ambiguities linger regarding the parameters for the host clusters \citep[e.g.,][and discussion therein]{ma95,ho03}. Indeed, the uncertainty led \citet{tu10} to bypass NGC 7790 when constructing the 2010 iteration of the Galactic Cepheid calibration.  Regarding M25, analyses are complicated by systematic offsets between relatively shallow observations published for the cluster, as well as its projection upon a rich background field.

New and existing multiband data are analyzed here with the aim of mitigating uncertainties tied to M25, NGC 7790, their key constituent Cepheids, and invariably the important Galactic Cepheid calibration \citep[][]{jl12,mo12,ng12b,gr13}.

\begin{figure*}[!t]
\begin{center}
\includegraphics[width=15.3cm]{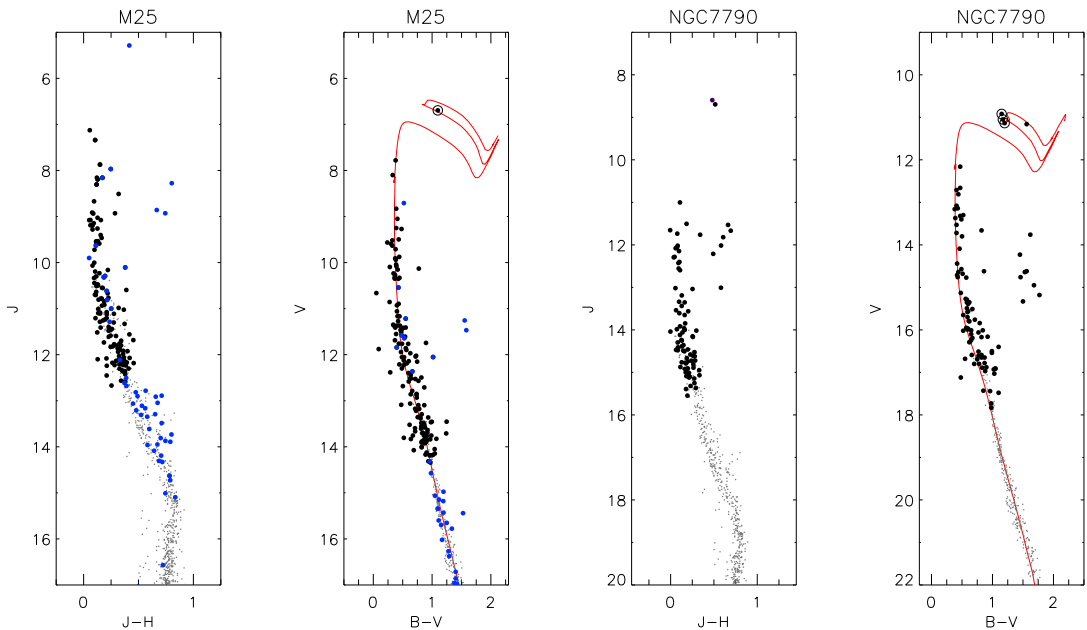} 
\caption{\small{Optical and near-infrared color-magnitude diagrams for M25 and NGC 7790.  For M25 the blue and black filled circles are X-ray and bright $U$-band selected sources, respectively.  Color cuts were made to reduce field star contamination. For NGC 7790 the black filled circles are 2MASS (near-infrared) and ARO (optical) field decontaminated \citep[][and references therein]{bb11} photometry.   Black dots in the color-magnitude diagrams represent the intrinsic relation \citep{ma11c} shifted to match the observations.  Optical photometry for the cluster Cepheids are conveyed by circled dots \citep{be00,be08}.}}
\label{fig-cmd}
\end{center}
\end{figure*}

\section{{\rm \footnotesize ANALYSIS}}
\subsection{{\rm \footnotesize M25}}
\label{s-M25}
Distinguishing cluster members from field stars is a principal challenge associated with establishing reliable parameters for open clusters. Late-type cluster dwarfs can prove difficult to distinguish from field stars that are of comparable brightness, hence the pertinence of X-ray data. \citet{ma12} used X-ray observations to identify low-mass stars associated with the Cepheid SU Cas, and its host cluster Alessi 95 \citep{tu12}. X-ray emitting stars associated with relatively young Cepheids \citep[$\tau\sim100$ Myr,][]{tu12a} are typically cooler than F5-F8~V, where convection becomes an important mode of energy transport in the stellar atmosphere. The X-rays stem from a hot coronal plasma driven by convection and the magnetic dynamo. X-ray emission drops with stellar age, since the angular momentum inherited from the accretion process dissipates (i.e., via stellar winds and magnetic braking) and the dynamo weakens. Consequently, relatively young stars associated with classical Cepheids can be separated from field stars along the same sight-line, since the latter are typically old slow-rotators that are X-ray quiet. X-rays are also useful for identifying companions to massive cluster stars \citep{ev11,ma12}.

XMM-Newton X-ray images (ObsID:040472, PI: Motch) of the field were examined to identify fainter cluster members (Fig.~\ref{fig-xmmfield}).   Sources were identified using SExtractor \citep{ba96}. The \textit{elongation} option was selected since sources on the XMM-Newton image were non-symmetric. A 2.5$\sigma$ criterion was chosen to reduce spurious detections. The X-ray targets were subsequently matched with newly acquired $UBV$ photometry and 2MASS near-infrared (AAA) photometry, and the resulting sample is displayed in Figs.~\ref{fig-xmmfield}, \ref{fig-cmd}. The new optical CCD photometry were obtained from the 1-m Henrietta Swope telescope (Las Campanas, Chile). The data were reduced following \citet{ca09}. Bright saturated stars were swapped with unpublished photoelectric photometry obtained from Las Campanas using the 0.6-m Helen Sawyer Hogg telescope.  Deeper near-infrared photometry was also obtained via the Osiris instrument on the 4-m SOAR telescope (Cerro Pach\'{o}n, Chile). The \citet{st87} DAOPHOT routines were employed to extract photometry from those images since the sources exhibited satisfactory stellar PSFs. The deeper near-infrared data revealed numerous field stars occupying the region of the color-magnitude diagram associated with cluster M-dwarfs.  2MASS observations constitute the bulk of the data displayed in the near-infrared color-magnitude diagram for M25 (Fig.~\ref{fig-cmd}), with photometry for several faint X-ray sources supplied by the SOAR observations. Deeper X-ray data are needed to fully exploit the SOAR data.

Field contamination was likewise minimized by selecting stars that exhibit bright $U$-band photometry.  Bright ($U$) early-type stars are comparatively rare in a field owing to a steep initial mass function and relatively short lifetimes. Moreover, background B-stars are faint given M25 is comparatively nearby. Most field stars that remained after imposing the $U$-band limit did not traverse the cluster sequence.  The combined X-ray and $U$-band selected stars populate the bulk of the cluster sequence (Fig.~\ref{fig-cmd}), thus facilitating the application of an isochrone.   Fewer X-ray stars were identified in the optical dataset relative to the near-infrared owing to the reduced field of view of the former (by contrast 2MASS is all-sky). 

The optical color excess was determined via the $UBV$ color-color diagram, and yielded $E_{B-V}=0.513\pm0.034 \sigma$ (Fig.~\ref{fig-ubvm25}). The intrinsic $UBV$ color-color relation of \citet[][and references therein]{tu89} was used, in tandem with the reddening line for the broader field determined by \citet{tu76}. B-stars were solely considered in the fit to avoid the Hyades anomaly \citep{tu79}. To within the uncertainties the cluster reddening matches the estimate cited for U Sgr by \citet[][$E_{B-V}=0.50\pm0.03$]{tu10}.  However, the cluster reddening is larger than the average color-excess cited for U Sgr by \citet[][and references therein, $E_{B-V}=0.42\pm0.02$]{lc07}.  The source of the reddening discrepancy is unknown, and differential extinction cannot be invoked since B-stars encompassing U Sgr share the mean cluster reddening.

The near-infrared reddening for M25 was inferred using spectral types cited for bright B-stars by \citet{wa61} and \citet{sk13}\footnote{Catalogue of Stellar Spectral Classifications.}, in conjunction with {\it JHK$_s$} intrinsic colors from \citet{sl09}. The resulting mean color excess is {\it E(J--H)} $=0.165\pm0.032 \sigma$. The \citet{wa61} spectral types are offset in temperature and luminosity class from those established by \citet{fe57} and \citet{wa57}. The latter, although comprehensive, often cite dwarf classifications for evolved B-stars. The evolved luminosity class determinations of \citet{wa61} are in better agreement with the color-magnitude diagram \citep[Fig.~\ref{fig-cmd}, see also][]{hc75}.  

With the reddening established, the isochrone and the intrinsic \citet{ma11c} relation were shifted along the ordinate of the color-magnitude diagram to match the observations.  The near-infrared empirical calibration derived by \citet{ma11c} was used to establish the cluster distance.  That calibration was constructed from revised Hipparcos parallaxes for nearby stars \citep{vle07}, and yields distances to 7 benchmark clusters (e.g., $\alpha$ Per) that are consistent with the \citet{vl09}\footnote{\citet{vl09} suggests that  Hipparcos parallaxes and the Hyades anomaly \citep[e.g., see \S 2 of][and Fig.~29 of \citealt{vl09}]{tu79} imply that traditional main-sequence fitting is flawed, as there exists an additional parameter affecting the luminosity (i.e., age). \citet{ma11c} provide an alternate interpretation to consider.} Hipparcos determinations. The relation is not tied to the Pleiades cluster, an important consideration given a dispute continues concerning its parameters \citep[][and references therein]{vl09,ma11c,dg12}.  A comprehensive analysis of B-stars in M25 by \citet{lu00} demonstrated that cluster stars exhibit near solar abundances.  U Sgr likewise displays near solar abundances \citep[][and references therein]{lc07,ks09}.  Metallicity effects in the near-infrared are negligible in this instance \citep{al96,ma11c}.  

\begin{figure*}[!t]
\begin{center}
\includegraphics[width=6.7cm]{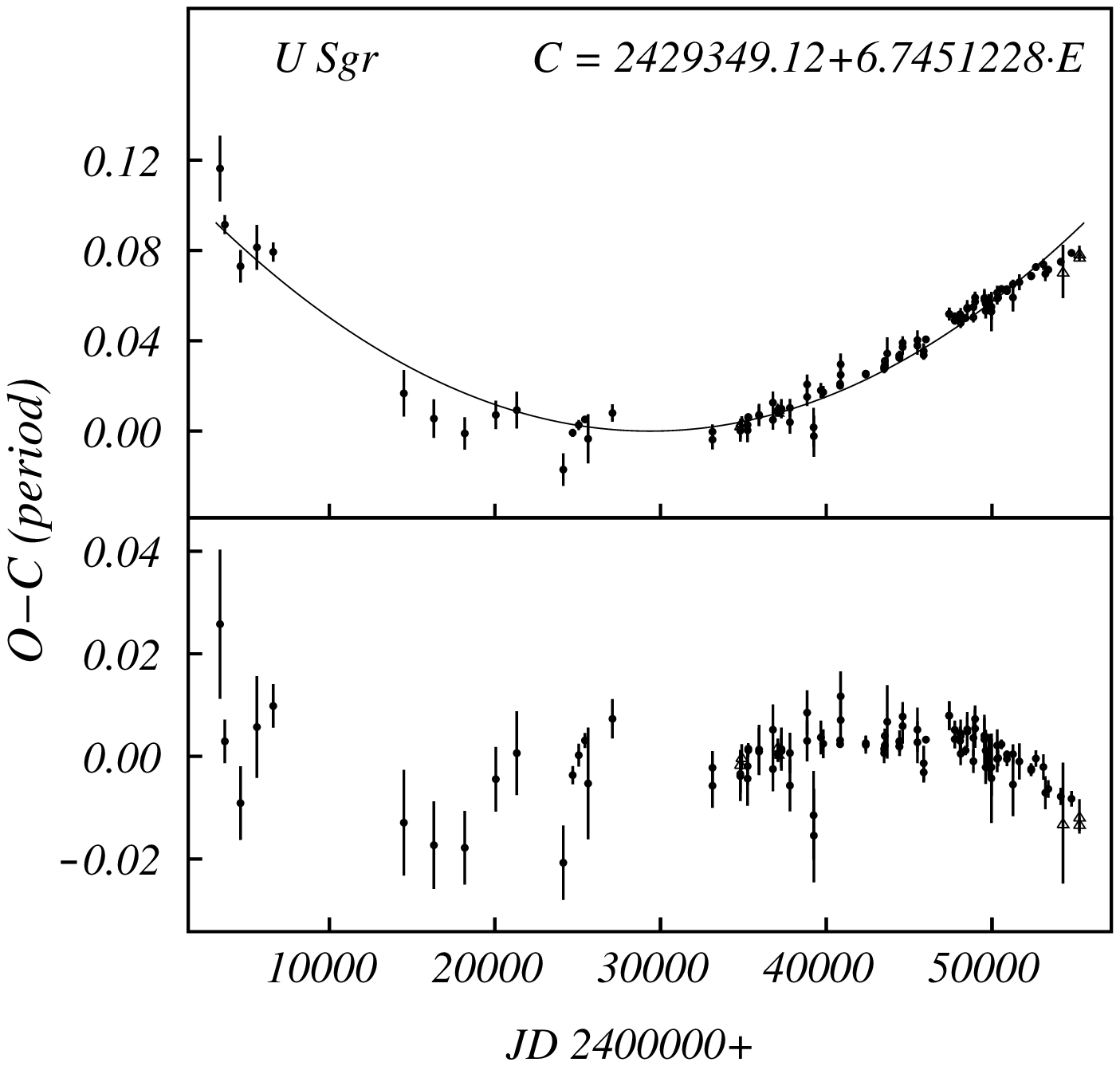} 
\includegraphics[width=6.7cm]{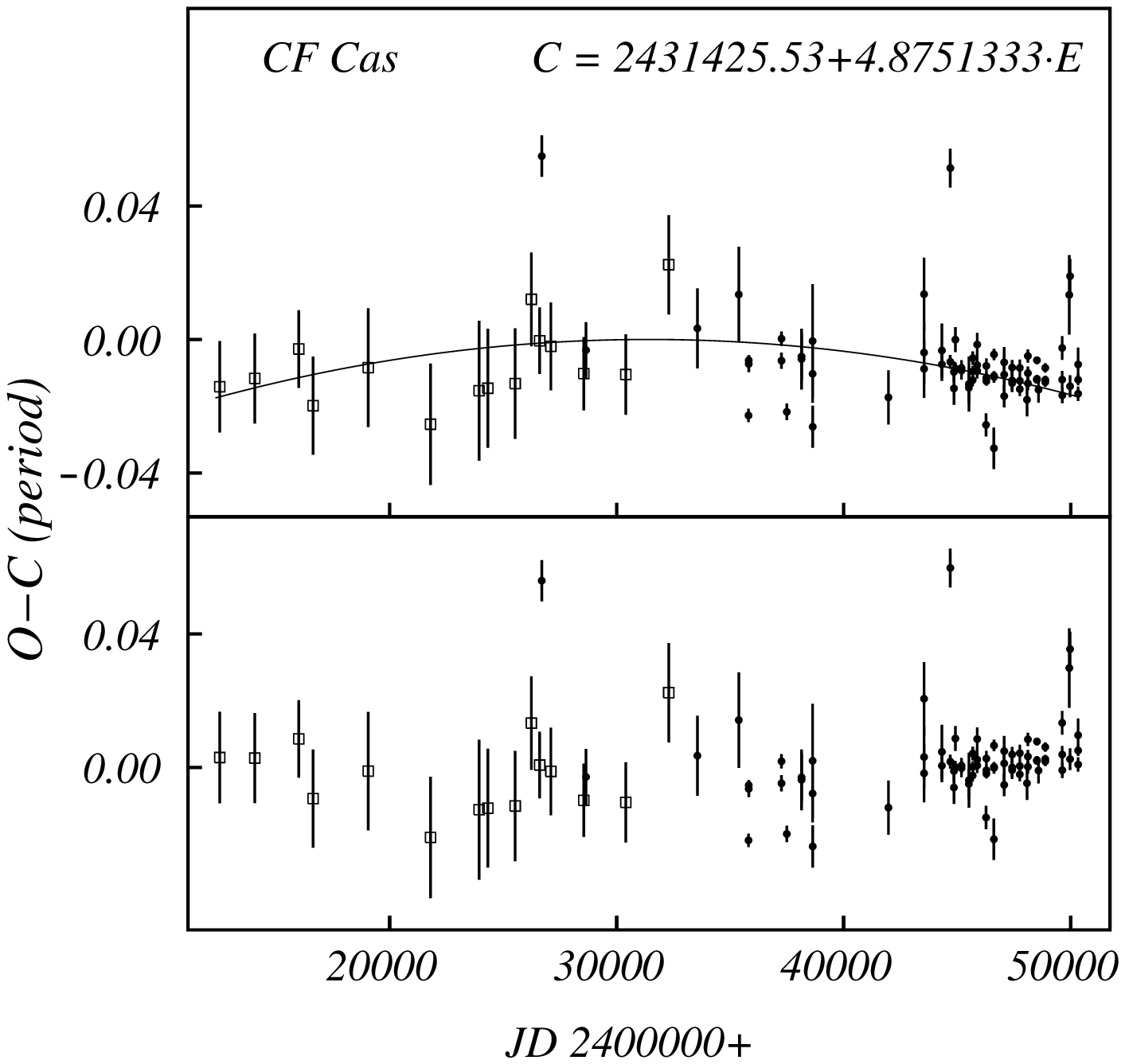} 
\caption{\small{Period evolution (O-C) diagrams for U Sgr ($dP/dt=+0.073$ s yr$^{-1}$) and CF Cas ($dP/dt=-0.072$ s yr$^{-1}$) facilitate the selection of an isochrone (Fig.~\ref{fig-cmd}). The new results imply that U Sgr is traversing the instability strip for the $3^{\rm rd}$ time and evolving toward the cool-end of the instability strip, whereas CF Cas may be a $2^{\rm nd}$ crosser venturing toward hotter temperatures.  The lower panels display the residuals from the polynomial fits applied.  \citet{st05} relays comprehensive information regarding period evolution (O-C) diagrams.}}
\label{fig-oc}
\end{center}
\end{figure*}

The resulting distance and age for M25 are $d=630\pm25$ pc and $\log{\tau}=7.88\pm0.10$, respectively.  The parameters were inferred from the near-infrared (the calibration) and optical data, respectively.   The latter dataset exhibits more reliable photometry for bright cluster members.  The distance determined for M25 agrees with a preliminary estimate deduced by \citet{tu10} \citep[see also][]{kh05}, yet is larger than the distance established for U Sgr from the infrared surface brightness technique \citep[][$d=579\pm6$ pc]{st11}.  The age was ascertained by matching the isochrone to evolved blue and red cluster stars, whereby a solar isochrone was selected \citep[Padova,][]{gi02,bo04}. The period evolution of U Sgr ($dP/dt=+0.073\pm0.010$ s yr$^{-1}$, Fig.~\ref{fig-oc}) was likewise considered in the aforementioned determination, since it implies that the Cepheid lies in the third crossing of the instability strip according to the \citet{tu06} framework \citep[see also][]{sz83,be85,ne12}. The duration of each instability strip crossing differs, whereby the first traversal is shortest \citep{be85}. The positive rate ($dP/dt$) implies U Sgr is evolving toward the cool side of the instability strip, provided the period evolution is primarily tracking density changes \citep{ed18}. However, recent and precise observations exhibit a superposed variation (Fig.~\ref{fig-oc}). The aforementioned conclusions stem from a comprehensive analysis featuring more than 100 years of U Sgr observations\footnote{An elaborate listing of references tied to the O-C analyses (Fig.~\ref{fig-oc}) is obtainable from the coauthor L.~Berdnikov.} (Fig.~\ref{fig-oc}).

An optical based distance was evaluated with the aim of supporting (to first-order) the near-infrared solution.  The analysis yielded $d=650$ pc, however, that result was tied to the canonical extinction law \citep[$R_V=A_V/E_{B-V}\sim3.1$,][]{tu76}. The ratio varies throughout the Galaxy, e.g., \citet{ca13} obtain $R_V\simeq3.85$ for the cluster Westerlund 2, while \citet{na13} derive $R_V=2.5$ for the Galactic bulge.  Majaess et al.~(2013-14, in preparation) mapped $R_V$ variations throughout the region and determined that $R_V\sim3.25$ (Fig.~\ref{fig-lrv}).\footnote{The Majaess et al.~(2013, in preparation) analysis utilized mid-infrared WISE data \citep{wr10} to determine $R_V$ via the color extrapolation method.}  That implies an optical-based distance for M25 ($d=640$ pc) that matches the near-infrared solution. Admittedly, obtaining new spectra to confirm existing classifications is desirable, and would allow for a separate estimate of $R_V$ \citep[e.g.,][]{we02}.

\subsection{{\rm \footnotesize NGC 7790}}
Reddening estimates for NGC 7790 exhibit a sizable spread and it was speculated that dust along the sight-line is peculiar \citep[][their Fig.~14 and references therein]{ho03}. However, the $UBV$ color-color diagram constructed by \citet{mm88} does not support that finding. It is unclear whether problems endemic to standardizing $U$-band photometry are the source of the discrepancy \citep{cc01,ma02,tu11}, or a Hyades-like anomaly is present.  

New spectroscopic observations were acquired to constrain the reddening. Six cluster B-stars were observed using the 1.8-m Plaskett telescope (Dominion Astrophysical Observatory (DAO), Victoria). The spectra were reduced via IRAF, and are shown in Fig.~\ref{fig-spectra}.  Spectral classifications and coordinates for the sample are listed in Table~\ref{table1}, the former being derived using the precepts highlighted by \citet{gc09}. The results are generally confirmed by fitting synthetic spectra to the observations \citep[see \S \ref{s-spectra}, and][]{mo07,Moni12}.  

Four additional B-stars featured in the \citet{sk13} catalog were included to evaluate a mean color-excess of: {\it E(J--H) =} $0.14\pm0.04\sigma$ and {\it E$_{B-V}$ =} $0.52\pm0.05\sigma$. The former was computed using 2MASS photometry and near-infrared intrinsic colors from \citet{sl09}, whereas the latter stemmed from new ARO\footnote{The Abbey Ridge observatory (ARO) is a remotely operated facility hosting a 0.3-m telescope, and is engaged in supernova, variable star, and cluster research \citep[e.g.,][]{lg11,ma12b}.} $BV$ photometry in conjunction with the intrinsic colors of \citet[][and references therein]{tu89}. To within the uncertainties the reddening for NGC 7790 is consistent with estimates cited previously for CF Cas \citep[][and references therein, $E_{B-V}=0.553$]{lc07,fo07}.

\begin{figure}[!t]
\begin{center}
\includegraphics[width=7cm]{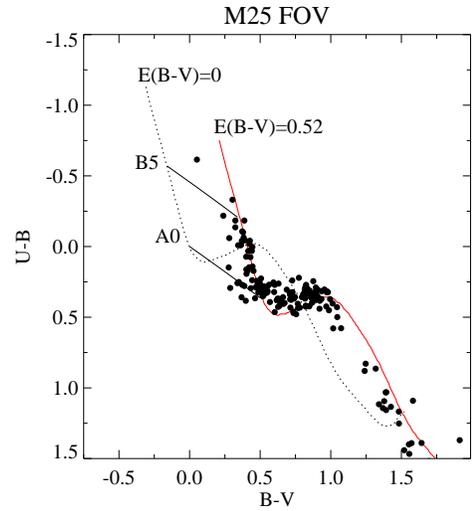} 
\caption{\small{New observations for stars in the field of M25 that exhibit bright $U$-band photometry.  The intrinsic relation and reddening law by \citet[][and references therein]{tu89} and \citet{tu76} were adopted, respectively.  A mismatch to the intrinsic relation appears just beyond A0, and may be indicative of a standardization offset or the Hyades anomaly.  A mean cluster reddening was determined ($E_{B-V}=0.513\pm0.034\sigma$) from the B-stars to avoid the aforementioned problem \citep{tu79}.}}
\label{fig-ubvm25}
\end{center}
\end{figure}

\begin{figure}[!t]
\begin{center}
\includegraphics[width=8.4cm]{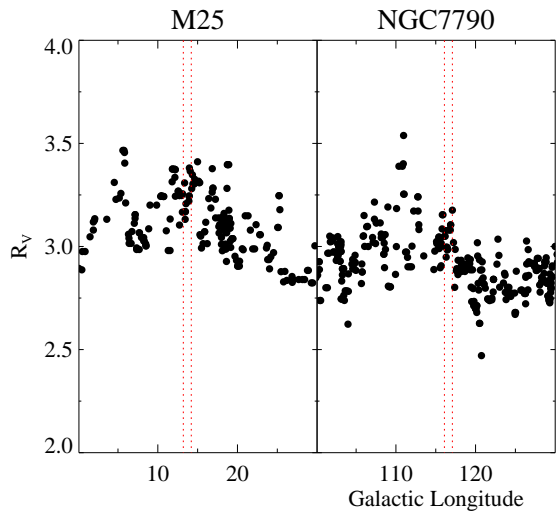} 
\caption{\small{\textit{Estimated} extinction law ($R_V=A_V/E_{B-V}$) variations binned as a function of Galactic longitude (Majaess et al.~2013-14, in preparation).  The broader sight-lines encompassing M25 ($\ell \sim 14 \degr$) and NGC 7790 ($\ell \sim 117 \degr$) exhibit larger and marginally smaller values of $R_V$ relative to the canonical extinction law \citep[$R_V\sim3.1$,][]{tu76}, respectively.}}
\label{fig-lrv}
\end{center}
\end{figure}

The optical CCD observations obtained at the ARO were standardized (Fig.~\ref{fig-cmd}) to unpublished photoelectric photometry obtained at Kitt Peak during August 1984 (the 0.9-m and 1.3-m telescopes were used in tandem with a 1P21 system). The ARO photometry was generated using DAOPHOT, and a search for variability was carried out via the VaST\footnote{Variability Search Toolkit (VaST): \url{http://scan.sai.msu.ru/vast/}} routine \citep{sl05}. No cluster variables were detected to sufficient precision other than the Cepheids. New $JHK_s$ photometry was obtained from the Observatoire Mont-M\'{e}gantic \citep{ar10} for the field hosting NGC 7790. However, the photometry could not be placed on the 2MASS system since non-linearities appeared near the faint-end of the standardization, thereby rendering the zero-point uncertain for any key data beyond the 2MASS limits. Those observations were thus omitted from the analysis.  

NGC 7790 is too distant to exhibit reliable proper motion or X-ray data that could facilitate the identification of cluster members. A field star decontamination algorithm was thus employed \citep[][and references therein]{bb11}, whereby the morphology displayed in the color-magnitude diagram by a field sample was removed from a diagram featuring field and cluster stars (of equal area). The results for NGC 7790 are shown in Fig.~\ref{fig-cmd}. 

The distance to NGC 7790 was established in the manner described for M25 (\S \ref{s-M25}). The near-infrared distance is $d=3.40\pm0.15$ kpc, which fits the optical-based solution (Fig.~\ref{fig-cmd}). The cluster distance and reddening are larger and marginally smaller, respectively, than estimates recently inferred from MegaCam images \citep[][$d\sim3176$ pc and $E_{B-V}=0.56$, see also his Table~2]{da12}. \citet{kh05} advocate cluster parameters of: $d=2944$ pc and $E(B-V)=0.53$.  The cluster age ($\log{\tau}=8.0\pm0.1$) was inferred by matching a solar isochrone to precise optical photometry for evolved members (the authors are unaware of complete near-infrared data for CEab Cas).  Here too an analysis of the Cepheids' period evolution was considered to facilitate the age determination. An examination of over 100 years of CF Cas observations implies that the Cepheid is in the second crossing ($dP/dt=-0.0721\pm0.014$ s yr$^{-1}$), given its low rate of period change \citep[the 2$^{\rm nd}$ crossing is predicted to be longest,][]{be85}. CEa and CEb Cas were analyzed by \citet{be90}, who noted that the Cepheids are in the third crossing. Photometry for the Cepheids \citep{be00,be08} support the aforementioned conclusions (i.e., CEa Cas is brighter than CF Cas), although uncertainties associated with the period-change and crossing-mode determined render the conclusion tentative.  However, the period analysis certainly indicates that none of the Cepheids are first-crossers (the crossing mode with the shortest duration, i.e., rapid period change).  

\begin{figure*}[!t]
\begin{center}
\includegraphics[width=14cm]{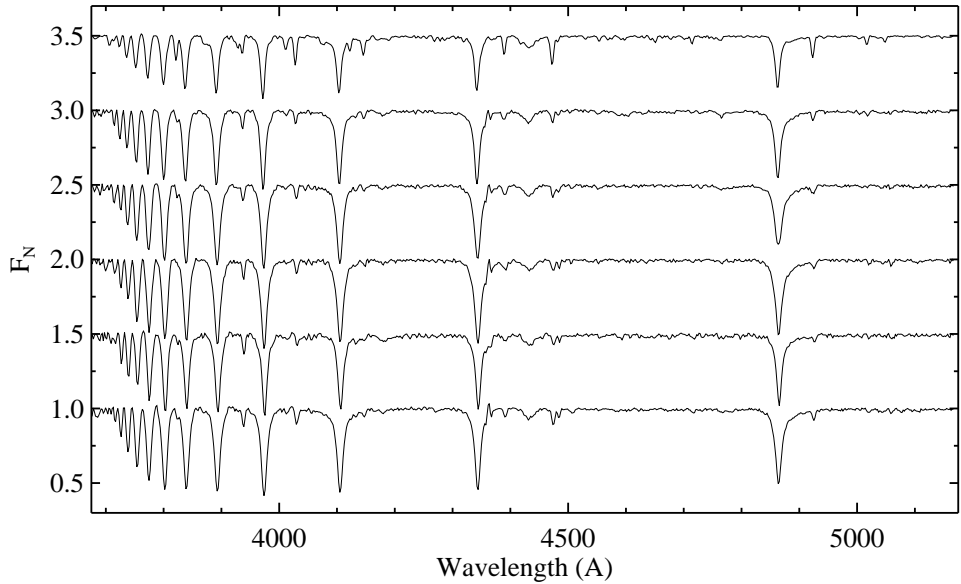} 
\caption{\small{New DAO spectra for B-stars in NGC 7790 were used in the determination of the mean reddening ($E_{B-V}=0.52\pm0.05\sigma$).  The spectra are listed in order of increasing ID.  Pertinent details regarding the stars are highlighted in Table~\ref{table1}.}}
\label{fig-spectra}
\end{center}
\end{figure*}

\begin{table*}
\caption{Spectroscopically Observed Stars Encompassing NGC 7790}
\label{table1}
\centering 
\begin{tabular}{cccccc}
\hline\hline
ID & RA/DEC (J2000) & SpT$^a$ & $V\sin{i}$ (km/s)$^b$  & T$_\mathrm{eff}$ ($\pm100$ K)$^b$  & $\log{g}$ ($\pm0.03$ dex)$^b$  \\
\hline
0$^c$ 	& 23:58:07.40	+61:11:44.9 &	B1V & 150 & 24500 & 4.11\\
1	& 23:58:28.10	+61:12:03.1	& B5IV  &	170 & 14700 & 3.63 \\
2	& 23:58:23.19	+61:12:25.0	& B7IV & 220 & 13400 & 3.72 \\
3	& 23:58:29.05	+61:12:41.4	& B8IV & 0 & 13400 & 3.98 \\
4	& 23:58:30.68	+61:12:18.7	& B8IV & 190 & 14000 & 3.60 \\
5	& 23:58:31.32	+61:12:49.8	& B6IV   & 190 & 13000 & 3.73 \\
\hline
\end{tabular}  \\
\begin{flushleft}
\begin{small}
a) Classified by visual inspection using the precepts outlined by \citet{gc09}.\\
b) Inferred from synthetic spectra fits to the observations (\S \ref{s-spectra}).  The statistical uncertainties produced by the fitting routine are cited, and may underestimate the combined uncertainty by $2-4\times$ (Napiwotzki 1999, priv.~comm.).\\
c) Spectroscopic parameters for evolved M25 stars are also presented by \citet[][their Table~2]{lu00}.
\end{small}
\end{flushleft}
\end{table*}

\subsubsection{Synthetic Spectra Fits}
\label{s-spectra}
Fundamental parameters for the DAO target stars (T$_\mathrm{eff}$, $\log g$, $V\sin{i}$) were estimated by fitting the Balmer and helium lines with synthetic spectra. Solar metallicity was assumed given that new abundance estimates derived for this study imply that CF Cas, CEa Cas, and CEb Cas have $[\rm{Fe/H}]=-0.08,-0.03,$ and $-0.01$ ($\sigma=0.15$ dex), respectively (Kovtyukh 2013, unpublished). A grid of model spectra (T$_\mathrm{eff}$=12\,000--32\,000~K and $\log g$=3.5--5.0~dex) was generated via Lemke's version\footnote{\url{http://a400.sternwarte.uni-erlangen.de/~ai26/linfit/linfor.html}} of the LINFOR program (developed originally by Holweger, Steffen, and Steenbock at Kiel University).  Local thermodynamic equilibrium (LTE) model atmospheres were computed with ATLAS9 \citep{Kurucz93}. Extensive details about the synthetic spectra fitting procedure are found in \citet{Moni12}. 

The stellar parameters were derived via a $\chi^2$ fit, as implemented in the {\it fitprof21} code developed by \citet{Bergeron92} and \citet{Saffer94}, and subsequently modified by \citet{Napiwotzki99}. The routine returns formal statistical uncertainties, and neglects uncertainties stemming from the normalization procedure, sky subtraction, and flat-fielding. Therefore the uncertainties cited are likely underestimated (Table~\ref{table1}). Balmer lines from H$\beta$ to H${12}$ were fitted simultaneously, with the exception of H$\epsilon$ (when blended with the Ca\,II~H line). The derived values of T$_\mathrm{eff}$ and $\log{g}$ are most sensitive to the hydrogen lines.  Four weak helium lines ($\lambda$4026, 4388, 4471, and 4922~\AA) were also matched, and those features generate an improved estimate for $V\sin{i}$ than broad Balmer lines. $V\sin{i}$ was an input quantity that was iteratively tweaked until a minimum $\chi^2$ was achieved. Admittedly, $V\sin{i}$ is not precisely determined at the resolution of the spectra obtained. Numerous factors may alter the effective line width (e.g., instrumental profile, variations of the true resolution owing to seeing), although rapid rotation endemic to such hot targets is the dominant line-broadening effect. A precise $V\sin{i}$ estimate would likewise require a zero-point calibration to standard stars \citep[e.g.,][]{RecioBlanco04,Monaco11}. The results presented in Table~\ref{table1} are thus deemed suggestive and may be affected by systematic offsets \citep[e.g., systematic offsets exist between parameters inferred from NLTE and LTE models,][]{ys13}.  

Results inferred from the synthetic spectra are consistent with the spectral types assigned from a visual inspection (to within the expected uncertainties), thus supporting the cluster parameters deduced.

\section{{\rm \footnotesize CONCLUSION}}
A new multiband X-ray, $UBVJHK_sW_{(1-4)}$, and spectroscopic investigation was carried out to establish precise cluster parameters (Table~\ref{table1}).  Parameters tabulated for M25 are $d=630\pm25$ pc and $\log{\tau}=7.88\pm0.10$, as deduced from isochrone fits to color-magnitude diagrams populated by X-ray and $U$-band selected members (Figs.~\ref{fig-xmmfield}, \ref{fig-cmd}). The age is partly associated with a comprehensive period-change analysis for U Sgr (Fig.~\ref{fig-oc}). The period evolution of U Sgr ($dP/dt=+0.073$ s yr$^{-1}$) implies the Cepheid is in the third crossing according to the \citet{tu06} framework, and evolving toward cooler temperatures. Dust along the line of sight to M25 is characterized by color excesses of $E(J-H)=0.165\pm0.032 \sigma$ and $E_{B-V}=0.513\pm0.034 \sigma$ (Fig.~\ref{fig-ubvm25}), the former being tied to the \citet{wa61} spectral types. The \citet{wa61} spectral types are offset from previously published designations, however, they match the evolved (IV-III) luminosity classifications expected for late-to-mid B-stars associated with intermediate-period Cepheids (Fig.~\ref{fig-cmd}). The Majaess et al.~(2013, in preparation) analysis implies that dust along the broader sight-line encompassing M25 follows a marginally larger extinction law ($R_V\simeq3.25$).  Adopting that value reduces the offset between the near-infrared and optical distance solutions for the cluster.

\begin{table*}
\caption{Cluster Parameters} 
\label{table2} 
\centering 
\begin{tabular}{lcccc} 
\hline\hline 
Cluster & $E(J-H)$ & $E(B-V)$ & d (pc) & $\log{\tau}$ \\ 
\hline 
M25 & $0.165\pm0.032\sigma$ & $0.513\pm0.034\sigma$& $630\pm25$ &  $7.88\pm0.10$ \\
NGC 7790 & $0.14\pm0.04\sigma$ & $0.52\pm0.05\sigma$ & $3400\pm150$ & $8.0\pm0.1$  \\
\hline 
\end{tabular}
\end{table*}

Parameters determined for NGC 7790 are $d=3.40\pm0.15$ kpc and $\log{\tau}=8.0\pm0.1$, where the results rely partly on the application of a field decontamination algorithm \citep[][and references therein]{bb11} since X-ray data are unavailable. The period evolution of the cluster's constituent Cepheids CF Cas and CEab Cas \citep{be90} were considered when deriving the cluster age. An elaborate period analysis for CF Cas ($dP/dt=-0.0721$ s yr$^{-1}$, Fig.~\ref{fig-oc}) indicates it is a $2^{\rm nd}$ crosser evolving toward the hot end of the instability strip. It was previously suggested that dust along the line of sight to NGC~7790 is anomalous, as implied by $UBV$ color-color diagrams constructed from various published datasets \citep[][their Fig.~14 and references therein]{ho03}. However, an inspection of the $UBV$ color-color diagram assembled by \citet{mm88} for cluster stars does not support that finding. Nonetheless, the impact of inhomogeneous or stellar parameter sensitive $U$-band photometry were reduced by utilizing unpublished $UBV$ photoelectric photometry, near-infrared photometry, and new DAO spectroscopic observations for several B-stars (Fig.~\ref{fig-spectra}). The observations imply color excesses of {\it E(J--H)} $=0.14\pm0.04\sigma$ and $E_{B-V}=0.52\pm0.05\sigma$, and rely on spectroscopic classifications assigned by visual inspection and generally confirmed via synthetic spectral fits (\S \ref{s-spectra}, Table~\ref{table1}).  $R_V$ for the broader region encompassing NGC 7790 is marginally smaller than the canonical $R_V\sim3.1$ extinction law (Fig.~\ref{fig-lrv}).  

Lastly, the Galactic Wesenheit ($VI_c$) function based on the \citet{tu10} and \citet{be07} calibrators was updated using the suite of new parameters established here, and elsewhere \citep[e.g., V340 Nor and QZ Nor,][]{ma13b}.  The resulting function is described by:
\begin{equation}
\label{eqn1}
W_{VI_c,0}=(-3.31\pm0.07) \times \log{P_0} - (2.56\pm0.06)
\end{equation}
$P_0$ is the pulsation period (fundamental mode), $W_{VI_c,0}$ is the absolute Wesenheit magnitude, and the apparent Wesenheit magnitude is $W_{VI_c}=V-2.55 \times (V-I_c)$.  The slope of the Galactic $W_{VI_c}$ function (Eqn.~\ref{eqn1}) matches that derived for LMC and SMC Cepheids \citep{ng09}, but is inconsistent with the slope established for distant solar extragalactic Cepheids associated with known supernova-host galaxies \citep{ma10,ri09}.  Continued research aimed at reducing uncertainties associated with the Cepheid calibration \citep[e.g.,][]{be07,gi13} \textit{and} the impact of crowding on remote extragalactic Cepheid photometry may inevitably help identify the source behind discrepant values of $H_0$ cited in the literature \citep[e.g.,][]{ri11,fr12,tr12,pl13}.

\subsection*{Acknowledgements}
\scriptsize{DM is grateful to the following individuals and consortia whose efforts, advice, or encouragement enabled the research: 2MASS (R. Cutri), HIP (F.~van Leeuwen), B.~Skiff, WEBDA (E.~Paunzen, J-C.~Mermilliod), DAML02 (W.~Dias), OMM (R.~LeMontagne, E.~Artigau), WISE, J.~Wampler, DAOPHOT (P.~Stetson), VaST (K.~Sokolovsky), VVV (D.~Minniti), CDS (F. Ochsenbein, T. Boch, P. Fernique), arXiv, and NASA ADS. LB thanks the Russian Foundation for Basic Research (project 13-02-00203). WG is grateful for support from the BASAL Centro de Astrofisica y Tecnologias Afines (CATA) PFB-06/2007.  DM and LB thank O. J. Knudsen (Aarhus Universitet) for providing access to a publication featuring pertinent U Sgr data (Nielsen A.~V., Medd.~Romer Observ.~Aarhus, 1954, Nr.~24, P.337-350).

This publication makes use of data products from the Wide-field Infrared Survey Explorer, which is a joint project of the University of California, Los Angeles, and the Jet Propulsion Laboratory/California Institute of Technology, funded by NASA; observations obtained with XMM-Newton, an ESA science mission with instruments and contributions directly funded by ESA Member States and NASA; observations obtained at the Southern Astrophysical Research (SOAR) telescope (program ID: CN2013A-157), which is a joint project of the Minist\'{e}rio da Ci\^{e}ncia, Tecnologia, e Inova\c{c}\~{a}o (MCTI) da Rep\'{u}blica Federativa do Brasil, the U.S. National Optical Astronomy Observatory (NOAO), the University of North Carolina at Chapel Hill (UNC), and Michigan State University (MSU).}

\end{document}